\documentclass[twocolumn,prl,aps,superscriptaddress,showpacs,amsmath,amssymb,floatfix]{revtex4-1}
\usepackage{color}
\usepackage{xcolor}
\usepackage{mathrsfs}
\usepackage{amsmath}
\usepackage{graphicx}
\usepackage{dcolumn}
\usepackage{bm}

\begin{document}

\title{Exact solutions for a type of electron pairing model with spin-orbit interactions and Zeeman coupling}%

\author{Jia Liu}
\affiliation{Department of Physics and Center of Theoretical and Computational Physics,The University of Hong Kong, Pokfulam Road,Hong Kong China}%

\author{Qiang Han}
\affiliation{Department of Physics, Renmin University, Beijing, China} %
\affiliation{Department of Physics and Center of Theoretical and Computational Physics,The University of Hong Kong, Pokfulam Road,Hong Kong China}%

\author{L.~B. Shao}
\affiliation{Department of Physics and Center of Theoretical and Computational Physics,The University of Hong Kong, Pokfulam Road,Hong Kong China}%

\author{Z.~D. Wang}
\email{zwang@hkucc.hku.hk} %
\affiliation{Department of Physics and Center of Theoretical and Computational Physics,The University of Hong Kong, Pokfulam Road,Hong Kong China}%

\date{\today}

\begin{abstract}
A type of electron pairing model with spin-orbit interactions or
Zeeman coupling is solved exactly in the framework of Richardson
ansatz. Based on the exact solutions for the case with spin-orbit
interactions, it is shown rigorously that the pairing symmetry is of
the $p$+i$p$-wave  regardless of the strength of pairing
interaction, 
as expected by the mean field theory. Intriguingly, how Majorana
fermions can emerge in the system is also elaborated. Exact
analytical results are illustrated for two simple systems
respectively with spin-orbit interactions and Zeeman coupling.
\end{abstract}

\maketitle


Recently, significant research attentions have been paid to various
physical systems with spin-orbit interactions, including Quantum
Spin Hall Effects\cite{hall_1, hall_2,Zhangsc_sci_2006}, topological
insulators~\cite{kpoint}, semiconductor heterostructures~\cite{kcross},
and a number of artificial systems like ultra-cold
atoms in optical lattices \cite{zhushiliang1,zhushiliang2}. In
particular, several important theoretical understandings have been
obtained for pairing electrons in the presence of spin orbit
interactions~\cite{kpoint, TI1}. Nevertheless, all of these
theoretical investigations on pairing systems have been conducted in
the framework of mean field theory, which is known to be a good
approximation merely for weak pairing interactions. Therefore, more
rigorous theoretical understandings or even exact solutions for these
electron pairing systems are highly appreciated, particularly for
strong pairing cases, even though it is extremely challenging to
find exact solutions of models for interacting many-electron
systems.  This is a central motivation of this work.

It is noted that Richardson obtained exact solutions of some pairing
models in the 1960s~\cite{Richard63}. As is known, Richardson's
exact solutions for pairing force models have played an important
role in the research of interacting many-particle
physics~\cite{RMP04}, including their connection with the well known
BCS model~\cite{ent}.

In this Letter, we first consider a type of electron pairing model
with spin-orbit interactions and solve it exactly in the framework
of Richardson ansatz. As an illustration, an analytical result is
derived for a very simple case.
Based on the exact solutions obtained, we show rigorously that the
pairing order parameter has always the $p$+i$p$-wave symmetry
regardless of the strength of pairing interactions, which recovers
an important conclusion deduced from the mean field theory.
Then, we address the same model with the Zeeman coupling
term~\cite{major}. Remarkably, we are also able to find an exact
solution in the presence of a pure Zeeman term  with the same
scenario. Exact analytical results are presented for a special
electron system.
Moreover,  we also elaborate how
Majorana fermions can emerge in the system.

Let us consider a pairing electron Hamiltonian with spin-orbit
interactions in a two dimensional lattice, which may be written as
\begin{equation}
H =H_{0}+ H_{int},\label{original_H}
\end{equation}
with
\begin{eqnarray*}
&H_{0}&=\sum_{{\bf{k}}}(c_{{\bf{k}}\uparrow}^{\dag},c_{{\bf{k}}\downarrow}^{\dag})
(\varepsilon_{{\bf{k}}}+\alpha{\mathbf{k}}\cdot{\bm{\sigma}})(c_{{\bf{k}}\uparrow},c_{{\bf{k}}\downarrow})^\text{T}, \\
&H_{int}&=-\sum_{{\bf{k}},\bf{k}'}V_{0}({\bf{k}},{\bf{k}}')c_{{\bf{k}}\uparrow}^{\dag}c_{{\bf{-k}}\downarrow}^{\dag}c_{{\bf{-k}}'\downarrow}c_{{\bf{k}}'\uparrow},
\end{eqnarray*}
where $\varepsilon_{{\bf{k}}}$ is the spin-independent single
electron energy, $c_{\bf{k}\uparrow(\downarrow)}^{\dag}$ and
$c_{\bf{k}\uparrow(\downarrow)}$
 are the creation and annihilation
operators of electrons, ${\bf{k}}=(k_{x},k_{y})$ is the wave vector
of the lattice~\cite{note1}, $\alpha$ is the effective strength of
spin-orbit interaction, and ${\bm{\sigma}}=(\sigma_{x},\sigma_{y})$
is the pauli matrices. We here consider an $s$-wave pairing
interaction, namely $V_{0}({\bf{k}},{\bf{k}}')=V_{0}>0$. Although
the above Hamiltonian has been studied recently under various mean
field approximations, to the best of our knowledge, it has not been
solved exactly. Motivated by this, we here want to find an exact
solution in the framework of  Richardson ansatz. We first
diagonalize the single-particle Hamiltonian by making the following
unitary transformation,
\begin{equation}
\begin{aligned}
    c_{{\bf{k}}\uparrow}&=&\frac{1}{\sqrt{2}}(a_{{\bf{k}},+} +e^{-i\theta({\bf{k}})}a_{{\bf{k}},-}), \\
    c_{{\bf{k}}\downarrow} &=& \frac{1}{\sqrt{2}}(e^{i\theta({\bf{k}})}a_{{\bf{k}},+}-a_{{\bf{k}},-}),\label{trans1}
\end{aligned}
\end{equation}
with $e^{i\theta({\bf{k}})}=(k_x+ik_y)/|\mathbf{k}|$~\cite{Wu} for
${\bf{k}}\neq 0$ and ~$e^{i\theta(0)}=1$ for ~${\bf{k}}=0$.
Physically, this unitary transformation corresponds to a local
spin-basis rotation to align the spin-direction along the wave
vector $\mathbf{k}$, which actually introduces an effective local
gauge field acting on electrons.  The Hamiltonian is now rewritten
as
\begin{equation}
\begin{aligned}
 H=& \sum_{{\bf{k}},s} \varepsilon_{{\bf{k}},s}
a_{{\bf{k}},s}^{\dag} a_{{\bf{k}},s}
-\frac{V_{0}}{4}\sum_{{\bf{k}}s,{\bf{k}}' s'} e^{-i
s\theta({\bf{k}})+is'\theta({\bf{k'}})}\\
&\times(A_{{\bf{k}},s}^{\dag}-\delta_{\mathbf{k},0}A_{0,0}^{\dag})
(A_{{\bf{k}}',s'}-\delta_{\mathbf{k}',0}A_{0,0}),
\label{pairingmodel}
\end{aligned}
\end{equation}
where the dispersion
$\varepsilon_{{\bf{k}},s}=\varepsilon_{{\bf{k}}} + s\alpha k$ with
$s=\pm 1$ denoting the two branches of the diagonalized
single-particle spectrum in the new basis.  Here, the pairing
operators are defined by
\begin{eqnarray}
& &A_{{\bf{k}},s}^{\dag}\equiv
a_{{\bf{k}},s}^{\dag}a_{-{\bf{k}},s}^{\dag} (s=\pm),\ \
A_{0,0}^{\dag}\equiv a_{0,+}^{\dag}a_{0,-}^{\dag}\ \ .
\label{operators}
\end{eqnarray}
In derivation of the above Eq.(\ref{pairingmodel}), we have employed a useful
relation ~$\theta(\bf{k})-\theta(-\bf{k})=\pm\pi$ for ${\bf{k}}\neq 0$. It is obviously seen from Eq.(\ref{operators}) that
$A_{{\bf{k}},s}^{\dag}
=0$ for ~${\bf{k}}=0$.
The above operators
satisfy the following commutation relations:
\begin{equation}
\begin{aligned}
& A_{{\bf{k}},s}^{\dag 2} =0, \\ %
 &[A_{{\bf{k}},s},A_{{\bf{k}}',s'}^{\dag} ] = \delta_{{\bf{k,k}}'}\delta_{s,s'}(1-2A_{{\bf{k}},s}^{\dag}A_{{\bf{k}},s}),\\
&[A_{{\bf{k}},s}^{\dag}A_{{\bf{k}},s},A_{{\bf{k}}',s'}^{\dag}]=\delta_{{\bf{k,k}}'}
\delta_{s,s'} A_{{\bf{k}},s}^{\dag} \label{comutation}
\end{aligned}
\end{equation}
for ${\bf{k}}$ or ${\bf{k}}'\neq 0$, and
\begin{eqnarray}
[A_{0,0},A_{0,0}^{\dag}]=
1-2A_{0,0}^{\dag}A_{0,0},\label{comutation2}
\end{eqnarray}
for  ${\bf{k}}={\bf{k}}'=0$. These relations
play a crucial role in solving this model exactly. Although the
pairing term in Eq.~(\ref{pairingmodel}) is $\mathbf{k}$-dependent,
we still make an ansatz in the same framework of Richardson's
pioneering work on a pairing model
~\cite{Richard63}.
In this framework, the eigenstates of
Hamiltonian (\ref{pairingmodel}) should take the product form as
\begin{equation}
|n,S_{+},S_{-}\rangle= \prod_{\mathbf{k}_{i}\in
S_{+}}a_{\mathbf{k}_{i},+}^{\dag} \prod_{\mathbf{k}_{j}\in
S_{-}}a_{\mathbf{k}_{j},-}^{\dag}
\prod_{\nu=1}^{n}B_{\nu}^{\dag}|0\rangle, \label{eigen}
\end{equation}
where
\begin{equation}
B_{\nu}^{\dag}=\sum_{\begin{subarray}{c}s,{\bf{k}}\in
P_s\\{\bf{{k}}\neq 0}\end{subarray}} \frac{e^{-i
s\theta({\bf{k}})}A_{\mathbf{k},s}^\dag}{2\varepsilon_{\mathbf{k},s}-E_{\nu}}+\frac{A_{0,0}^\dag}{2\varepsilon_{0}
-E_{\nu}}.
\label{ansatz1}
\end{equation}
Here $S_\pm$ denotes the set of singly occupied levels (namely
blocked levels) of the $\pm$ branch with cardinality $m_\pm$, while
$P_\pm$ the set of levels with the blocked ones excluded. The state
vector defined in Eq.~(\ref{eigen}) describes an eigenstate of
$N_{e}=m_{+}+m_{-}+2n$ electrons with $n$ as the number of electron
pairs. $E_\nu$'s in Eq.~(\ref{ansatz1}) are the parameters to be
determined by $n$ coupled algebraic equations to be given in the
following.

Solving the Schr\"{o}dinger equation associated with Hamiltonian (\ref{pairingmodel}) and the
eigenvector in Eq.~(\ref{eigen}) more tediously,
we obtain the equations for the present  two-branch electron system,
\begin{equation}
\begin{aligned}
1-\sum_{\begin{subarray}{c} s,\mathbf{k}\in P_s\\{\bf{{k}}\neq
0}\end{subarray}}\frac{V_{0}/2}{2\varepsilon_{\mathbf{k},s}-E_\nu}&-\frac{V_0}{2\varepsilon_{0}
-E_{\nu}}
+\sum_{\mu\neq\nu}^{n}\frac{2V_{0}}{E_\mu-E_\nu}=0,\label{FinalEquation}
\end{aligned}
\end{equation}
where $\nu=1,2,\ldots,n$.
The corresponding eigen-energy
is given by
\begin{equation}
E(n, m_{+}, m_{-})=\sum_{{\bf{k}}\in
S_+}\varepsilon_{{\bf{k}},+}+\sum_{{\bf{k}}\in
S_-}\varepsilon_{{\bf{k}},-}+\sum_{\nu=1}^{n} E_\nu.
\label{TotalEnergy}
\end{equation}
Remarkably, here we have demonstrated the pairing model of
Eq.(\ref{pairingmodel}) to be an integrable
problem~\cite{integrable}, making such pairing model more promising
and useful. The set of Eq.~(\ref{FinalEquation}) is quite similar to
Richardson's one. In particular, when $\alpha=0$, the two branches
are degenerate and Eq.~(\ref{FinalEquation}) recovers the usual
Richardson's equation~\cite{Richard63,E2,E3}.
It has been shown by Gaudin that Eq.~(\ref{FinalEquation}) has a continuum limit form
in the thermodynamic limit\cite{Gaudin,limit}.

Note that there are normally two kinds of spin-orbit interactions:
one takes the form of $\mathbf{k} \cdot \bm{\sigma}$ as in
Eq.~(\ref{original_H}) \cite{kpoint} with the exact solution being
given above, while the other has the form $(\bm{\sigma}\times
\mathbf{k})\cdot\hat{z}$ \cite{kcross,pwave}. If the spin-orbit
interaction in Eq.~(\ref{original_H}) is changed to the second form,
one can replace $\theta(\bf{k})$ in Eqs. (\ref{trans1}),
(\ref{pairingmodel}) and (\ref{ansatz1}) by
$\theta'(\mathbf{k})=\theta(\mathbf{k})-\pi/2$ and accordingly the
pairing model with $(\bm{\sigma}\times \mathbf{k})\cdot\hat{z}$-type
spin-orbit interaction is exactly solvable as well.

Although it is still rather challenging to solve
Eq.~(\ref{FinalEquation}) even numerically, the computational
loading is significantly reduced in comparison with the numerical
exact-diagonalization. In terms of this exact solution for the
system described by Hamiltonian (\ref{pairingmodel}), we are able to
evaluate some quantities exactly 
and obtain relevant rigorous results, which are very helpful for
validating or invalidating the related results based on the usual
mean field framework.

As an important example, we use Eq.~(\ref{eigen}) to calculate
exactly the following dimensionless order parameter,
\begin{equation}
\triangle_{{\mathbf{k}},s}= \frac{\langle
0,0,n-1|a_{{\bf{-k}},s}a_{{\bf{k}},s}|n,0,0\rangle}{\sqrt{C_n C_{n-1}}} = e^{-is\theta(\mathbf{k})} \triangle_{{\bf{k}},s}^0,
\label{order_para}
\end{equation}
where $C_n=\langle 0,0,n|n,0,0\rangle$ and
\begin{equation}
\triangle_{{\bf{k}},s}^{0} = \frac{1}{\sqrt{C_n C_{n-1}}} \sum_{\nu=1}^{n}
\frac{\sum_{\{j_i\}}^{\forall k_{j_i}\neq k} g_\nu^{(n)} g_n^{(n-1)*}}
{\varepsilon_k-E_\nu^{(n)}},
\end{equation}
with  
$$
g_\nu^{(n)} = \sum_P \prod_{\mu=1}^{\nu-1}\frac{1}{\varepsilon_{k_{j_\mu}}-E_{P\mu}^{(n)}}
\prod_{\mu=\nu+1}^{n}\frac{1}{\varepsilon_{k_{j_{\mu-1}}}-E_{P\mu}^{(n)}}.
$$
Here $k$ denotes $(\mathbf{k},s)$, the superscript $(n)$ corresponds
to the $n$-pair state, and $P$ means the permutation of the
corresponding terms. In the weak interaction limit
$(V_{0}\rightarrow 0)$, $E_\nu$'s are all real, so that
$\triangle_{{\bf{k}},s}^{0}$  is  real as well. It is clearly seen
that $\triangle_{{\bf{k}},s}$ has the $p_{x}+ip_{y}$ pairing
symmetry. Notably, even when the solutions $E_\nu$ are complex
numbers, we can show from Eq.(12) that $\triangle_{{\bf{k}},s}^{0}$
is still real, because the complex solutions of
Eq.~(\ref{FinalEquation}) appear in the form of conjugate pairs.
This finding for the pairing symmetry justifies a result expected by
the mean filed theory in the weak interaction
limit~\cite{kpoint,kcross}. In addition, we can also evaluate the
off-diagonal long range order,
\begin{equation}
\begin{aligned}
O(\mathbf{k}^\prime s^\prime,\mathbf{k}s) &\equiv C_n^{-1} \langle 0,0,n| A_{\mathbf{k}^\prime s^\prime}^\dagger A_{\mathbf{k}s} |n,0,0 \rangle, \\
& = e^{i[s^\prime\theta(\mathbf{k}^\prime)-s\theta(\mathbf{k})]} G(k^\prime,k),
\end{aligned}
\end{equation}
with $ G(k^\prime,k)=C_n^{-1}\sum_{\mu,\nu=1}^{n}
\frac{\sum_{\{j_i\}}^{\forall k_{j_i}\neq k, \neq k^\prime}
g_\mu^{(n)}
g_\nu^{(n)*}}{[\varepsilon_{k^\prime}-E_\nu^{(n)}][\varepsilon_{k}-E_\mu^{(n)}]},
$ which is always real and approaches $\Delta_{\mathbf{k}s}^0
\Delta_{\mathbf{k}^\prime s^\prime}^{0}$ in the thermodynamic limit
as expected.



For illustration of the complex solutions of Eq.~(\ref{FinalEquation}), let us look into
a toy model with four single-particle states (${\mathbf{k}\uparrow}$,
${-\mathbf{k}\downarrow}$, ${-\mathbf{k}\uparrow}$, ${\mathbf{k}\downarrow}$), which
accommodate four electrons. In the presence of the spin-orbit
interaction, the degenerate states are split into two groups with
$\varepsilon_{\mathbf{k},+}=\varepsilon_{-\mathbf{k},+}=1$ (as the energy unit) and
$\varepsilon_{\mathbf{k},-}=\varepsilon_{-\mathbf{k},-}=-1$. The set of Richardson equations
for this model are two coupled equations,
which are solved analytically to obtain,
\begin{equation}E_{1,2}=-V_0\pm
\sqrt{4-V_0^2}.
\end{equation}
One can find readily that $V_{0}=2$ is a critical value for
$E_{1,2}$ to become complex, while the ground-state energy $(E_{1}+E_{2})=-2V_0$ is always
real.

Next we turn to consider the Zeeman term \cite{major} induced by an
external magnetic field $\mathbf{B}=(B_x,B_y,B_z)$, which reads
$\hat{H}_Z
=\sum_{{\bf{k}}}(c_{{\bf{k}}\uparrow}^{\dag},c_{{\bf{k}}\downarrow}^{\dag})\mathbf{B}\cdot\bm{\sigma}(c_{{\bf{k}}\uparrow},c_{{\bf{k}}\downarrow})^\text{T}$
and is added to
 Hamiltonian (\ref{original_H}). We now make another transformation as
\begin{equation}
\begin{aligned}
    c_{{\bf{k}}\uparrow}&=& \cos\varphi_{\mathbf{k}} a_{{\bf{k}},+} + \sin\varphi_{\mathbf{k}} e^{-i\tilde{\theta}({\bf{k}})} a_{{\bf{k}},-}, \\
    c_{{\bf{k}}\downarrow} &=&
    \sin\varphi_{\mathbf{k}}e^{i\tilde{\theta}({\bf{k}})}a_{{\bf{k}},+} - \cos\varphi_{\mathbf{k}} a_{{\bf{k}},-},\label{trans2}
\end{aligned}
\end{equation}
where
\begin{equation}
\begin{aligned}
& \tan(2\varphi_{\mathbf{k}})=\eta_\mathbf{k}/B_z, & \\
& e^{i\tilde{\theta}(\mathbf{k})}=[(B_x+\alpha k_x)+i(B_y+\alpha
k_y)]/\eta_\mathbf{k}, &  \\
& \eta_\mathbf{k}=\sqrt{(B_x+\alpha k_x)^{2}+(B_y+\alpha k_y)^{2}}.
&
\end{aligned}
\label{ren}
\end{equation}
The single-particle spectrum still
has two branches with $
\varepsilon_{\mathbf{k},s}=\varepsilon_{{\mathbf{k}}}+s\sqrt{\eta_\mathbf{k}^2+B_z^{2}}.
$ In addition to the operators in Eq.(\ref{operators}), we also need
new operators defined as
\begin{eqnarray}
A_{{\bf{k}},0}^{\dag}\equiv
a^{\dag}_{{\bf{k}},+}a^{\dag}_{-{\bf{k}},-}, \
A_{{\bf{k}},0}\equiv
a_{-{\bf{k}},-}a_{{\bf{k}},+}.
\end{eqnarray}
Under this transformation, the total Hamiltonian with Zeeman term
can be rewritten as
\begin{equation}
\begin{aligned}
 H&= \sum_{{\bf{k}},s=\pm} \varepsilon_{{\bf{k}},s}
a_{{\bf{k}},s}^{\dag} a_{{\bf{k}},s} \\
&- V_{0}\sum_{{\bf{k}}s,{\bf{k}}' s'}e^{-i
s\tilde{\theta}(-s{\bf{k}})+is'\tilde{\theta}(-s'{\bf{k'}})}
\lambda^{\ast}_{s}(\mathbf{k})\lambda_{s'}(\mathbf{k}')A_{{\bf{k}},s}^{\dag}A_{{\bf{k}}',s'},
\label{s_H_M}
\end{aligned}
\end{equation}
where
\begin{equation*}
\begin{aligned}
&\lambda_{s}(\mathbf{k})\equiv s \cos
\varphi_{s\mathbf{k}}\sin\varphi_{s(-\mathbf{k})}
\\
&\lambda_{0}(\mathbf{k})\equiv
-\left(\cos\varphi_\mathbf{k}\cos\varphi_{-\mathbf{k}}+\sin\varphi_\mathbf{k}\sin\varphi_{-\mathbf{k}}
e^{i\tilde{\theta}(\mathbf{k})-i\tilde{\theta}(-\mathbf{k})}\right)
\end{aligned}
\end{equation*}
and $s,s'=0,\pm 1$ in the second summation of Eq.(\ref{s_H_M}) .

Generally, because $[A_{{\bf{k}},\pm}^{\dag},A_{{\bf{k}'},0}]\neq 0$
for ${\bf{k}}\neq 0$ and $\lambda_{s}(\mathbf{k})$ is
{\bf{k}}-dependent, it is hard to find an exact solution of
Hamiltonian (\ref{s_H_M}) by adopting a similar ansatz used above.
Nevertheless, Hamiltonian (\ref{s_H_M}) can still be solved exactly
for some special but relevant cases. When the external magnetic
field ~$\mathbf{B}=0$, we have
~$\varphi_{\mathbf{k}}=\varphi_{-\mathbf{k}}=\sqrt{2}/2$ and
$\tilde{\theta}(\mathbf{k})=\theta(\mathbf{k})$, so that
$\lambda_{s}(\mathbf{k})=s/2$ ($s=\pm$), $\lambda_{0}(\mathbf{k})=0 $
~$(\mathbf{k}\neq 0)$ and ~$\lambda_{0}(0)=-1$. In this case,
Hamiltonian (\ref{s_H_M}) reduces to Eq.~(\ref{pairingmodel}).

On the other hand,
when $\alpha=0$, $B_x=B_y=0$, and $B_z\neq 0$, 
only the Zeeman term is present. For this case,
$\lambda_{s}(\mathbf{k})=0$ ($s=\pm$) and $\lambda_{0}(\mathbf{k})=-1$.
 Since we have
\begin{equation}
[ \sum_{s=\pm} \varepsilon_{{\mathbf{k}},s} a_{{\bf{k}},s}^{\dag}
a_{{\bf{k}},s},A_{{\bf{k}},0}^{\dag}]=(\varepsilon_{{\bf{k}},+}+\varepsilon_{{\bf{k}},-})A_{{\bf{k}},0}^{\dag},
\end{equation}
we can take another ansatz
\begin{equation}
C_\nu^{\dag}=\sum_{{\bf{k}}}
\frac{A_{\mathbf{k},0}^\dag}{\varepsilon_{\mathbf{k},+}+\varepsilon_{\mathbf{k},-}-E_\nu}
\end{equation}
to replace ~$B_\nu^{\dag}$ in Eq.~(\ref{eigen}). Solving the Schr\"{o}dinger equation with
Hamiltonian (\ref{s_H_M}) and the corresponding eigenvector, we obtain the
equations that the parameters $E_\nu$'s satisfy,
\begin{equation}
1-\sum_{{\bf{k}}}\frac{V_{0}}{\varepsilon_{\mathbf{k},+}+\varepsilon_{\mathbf{k},-}-E_\nu}+\sum_{\mu\neq\nu}^{n}\frac{2V_{0}}{E_\mu-E_\nu}=0,\label{zeeman_Eqution}
\end{equation}
where $\nu=1,2,\ldots,n$. The expression of the eigenenergy of the
whole system is the same as Eq.(\ref{TotalEnergy}).
Eq.~(\ref{zeeman_Eqution}) implies that even when the single-particle
energies of electrons are spin-dependent, the Hamiltonian is
still exactly solvable. With the solutions of
Eq.(\ref{zeeman_Eqution}), we can similarly evaluate the dimensionless order
parameter 
for this
system with Eq.(\ref{order_para}). 
Now the order parameter $\triangle_{{\bf{k}},0}=\triangle_{{\bf{k}},0}^{0}$, which has just the usual $s$-wave symmetry with the same reason being mentioned before regardless of the strength of pairing  interactions.

As an interesting example, we also address a special case, where all
$N_{e}$ electrons are on the Fermi
surface $k=k_\text{F}$ in 
Eq.(\ref{zeeman_Eqution}), which is an approximation for the
considered system as many physical phenomena are only closely
related to the electrons near the Fermi surface. In this case, we
are able to obtain the total energy of the system without having all
$E_{\nu}$ to be solved from Eq.~(\ref{zeeman_Eqution}).  Supposing
the degeneracy of Fermi level to be $\Omega$ and considering that
$2\varepsilon_{{\bf{k}_\text{F}}}=\varepsilon_{\mathbf{k},+}+\varepsilon_{\mathbf{k},-}$,
we multiply $2\varepsilon_{{\mathbf{k}_\text{F}}}-E_\nu$ on
Eq.(\ref{zeeman_Eqution}) and take the sum of all $n$ equations to
obtain one equation. From this equation, we get the total pairing
energy $E_{P}$ of the pairing state $|n,0,0\rangle$ as
\begin{equation}
\begin{aligned}
E_{P}=2n\varepsilon_{{\mathbf{k}_\text{F}}}-V_{0}\Omega
n+2V_{0}\sum_{\nu}^{n}\sum_{\mu\neq\nu}^{n}\frac{2\varepsilon_{{\mathbf{k}_\text{F}}}-E_{\nu}}{E_\mu-E_\nu}.
\label{example}
\end{aligned}
\end{equation}
After partitioning the summation term into two parts with the sum
indices $\nu$, $\mu$ in one of them being interchanged, the total energy
of pairing system is given by
\begin{equation}
\begin{aligned}
E_{P}&=&2n\varepsilon_{{\mathbf{k}_\text{F}}}-V_{0}n(\Omega-n+1).
\end{aligned}
\end{equation}

Note that
$\varepsilon_{{\bf{k}},-}=(\varepsilon_{{\mathbf{k}_\text{F}}}-B_{z})<\varepsilon_{{\bf{k}},+}
=(\varepsilon_{{\mathbf{k}_\text{F}}}+B_{z})$,
single electrons prefer to occupy the $S_{-}$ set. Therefore, the
total energy of $|n,0,S_{-}\rangle$ is found to be
\begin{equation*}
\begin{aligned}
E=(N_{e}-2n)(\varepsilon_{{\mathbf{k}_\text{F}}}-B_{z})+2n\varepsilon_{{\mathbf{k}_\text{F}}}-V_{0}n(\Omega'-n+1),
\end{aligned}
\end{equation*}
where $\Omega'=\Omega-(N_{e}-2n)$ in consideration of that
$(N_{e}-2n)$ levels are blocked by single electrons.
Thus the condensation energy $\triangle{E}=E-E_{0}$ with $E_{0}$ as the no-pair
state energy
 is given by
\begin{equation}
\begin{aligned}
\triangle{E} =-V_{0}n^{2}+n[2B_{z}-V_{0}(\Omega-N_{e}+1)].
\end{aligned}
\end{equation}
Combining the condition $\triangle{E}\le 0$ with 
the  requirement of minimum $E$, we can readily find  a critical
value $B_c=V_{0}(\Omega-N_{e}/2+1)/2$. When $B_z<B_c$, the system is
in the pairing state, otherwise the ferromagnetic state.

We now attempt to elaborate how Majorana fermions(MF) can
emerge in the system described by Hamiltonian (1) based on our exact
solution. To capture the essential physics but without loss of
generality, we consider that $2n$ electrons occupy the states in a
narrow ribbon around $\varepsilon(\mathbf{k}_{F})$ and confined in
an annular region $r_{0}<r<R_{0}$.
In the continuum limit and from $H_0$, in addition to the bulk states $\varepsilon_{{\bf{k}},s}$
one can also find  the inner  and outer edge states with the energies $E_{in}=\varepsilon(\mathbf{k}_{F})+\alpha L_{z}/r_{0}$  and $E_{out}=\varepsilon(\mathbf{k}_{F})-\alpha L_{z}/R_{0}$, where $L_{z}$ is the angular momentum.
In the presence of the pairing interaction, the zero modes with
$L_{z}=0$ survive due to the topological protection~\cite{major} and
they could be occupied by pairs of MFs:
$a_{MF}(m{\mathbf{k}_{F}})=(\gamma_{1m}-i\gamma_{2m})/2$ ($m=\pm$)
with $\gamma_{im}=\gamma^{\dag}_{im}$ the MF-operators, while the occupied bulk
states are in the pairing states described by
Eq.(\ref{eigen}) with a lower energy. If an occupied
pairing state in the branch $\varepsilon_{{\bf{k}},+}$ is lowered
by the pairing energy to touch the edge state energy level
$\varepsilon(\mathbf{k}_{F})=E_{MF}$ with $E_{MF}$ the occupation energy for one pair of $m$-MFs, i.e.,
$E(n,0,0)=E(n-1,2,0)=E(n-1,0,0)+2E_{MF}$ in terms of
Eqs.~(\ref{eigen}) and (\ref{TotalEnergy}),
the MFs may emerge as gapless excitations. This condition also shows the degeneracy of occupation and
vacuum of MF states,  which the nonabelian statistics originates
from. Note that the probability amplitude for the emergence of a pair of
$m$-MFs is proportional to
$\langle0,1_{\mathbf{k}_{F}}+1_{-\mathbf{k}_{F}},n-1|
\gamma^{\dag}_{1m}\gamma^{\dag}_{2m}|n-1,1_{\mathbf{k}_{F}}+1_{-\mathbf{k}_{F}},0\rangle
\neq 0$
for the present system.
The above analysis asserts some important results
of the mean field theory~\cite{read,stone}.

In summary, by making a spin-rotation unitary transformation and in
the framework of Richardson ansatz, we  have
found for the first time that a class of electron pairing model with two kinds of
spin-orbit interactions is exactly solvable, which is closely relevant to recent
research hot spots on topological superconductors and Dirac
fermions. More importantly, based on the exact solution, we have rigorously shown that the
pairing symmetry is of the $p$+i$p$-wave
regardless of the strength of pairing interactions. Intriguingly, we have also elaborated how
Majorana fermions can emerge in the system.
Moreover, we have addressed a system
with the Zeeman term included and presented an exact solution as
well. Exact analytical results have been illustrated for two simple
examples. 
Finally, we wish to pinpoint that the present exact solutions for
the mentioned pairing systems may shed light on profound
understandings of topological superfluids.

We would like to thank  L.~A.~Wu, Y.~C.~He ,Y. Chen and Y. Li for
helpful discussions. This work was supported by the RGC of Hong Kong
(Nos. HKU7044/08P and HKU7055/09P) and a CRF of Hong Kong.


\end{document}